\documentstyle[twoside,fleqn]{article}
\begin{document}
\catcode`\@=11 
\oddsidemargin  -4mm              
\evensidemargin  4mm              

\topmargin      0mm              
\headheight     13mm              
\headsep        21pt              
\footskip       30pt              

\textheight 212mm                 
\textwidth 160mm                  

\columnsep 10mm                   
\columnseprule 0pt                

\parskip 0pt                      
\parindent 1em                    

\newdimen\@bls                    
\@bls=\baselineskip               
\advance\@bls -1ex                
\newdimen\@eps                    %
\@eps=0.0001pt                    

\def\section{\@startsection{section}{1}{\z@}
  {1.5\@bls plus 0.5\@bls}{1\@bls}{\normalsize\bf}}
\def\subsection{\@startsection{subsection}{2}{\z@}
  {1\@bls plus 0.25\@bls}{\@eps}{\normalsize\bf}}
\def\subsubsection{\@startsection{subsubsection}{3}{\z@}
  {1\@bls plus 0.25\@bls}{\@eps}{\normalsize\bf}}
\def\paragraph{\@startsection{paragraph}{4}{\parindent}
  {1\@bls plus 0.25\@bls}{0.5em}{\normalsize\bf}}
\def\subparagraph{\@startsection{subparagraph}{4}{\parindent}
  {1\@bls plus 0.25\@bls}{0.5em}{\normalsize\bf}}

\def\@sect#1#2#3#4#5#6[#7]#8{\ifnum #2>\c@secnumdepth
  \def\@svsec{}\else
  \refstepcounter{#1}\edef\@svsec{\csname the#1\endcsname.\hskip0.5em}\fi
  \@tempskipa #5\relax
  \ifdim \@tempskipa>\z@
    \begingroup
      #6\relax
      \@hangfrom{\hskip #3\relax\@svsec}{\interlinepenalty \@M #8\par}%
    \endgroup
    \csname #1mark\endcsname{#7}\addcontentsline
      {toc}{#1}{\ifnum #2>\c@secnumdepth \else
        \protect\numberline{\csname the#1\endcsname}\fi #7}%
  \else
    \def\@svsechd{#6\hskip #3\@svsec #8\csname #1mark\endcsname
      {#7}\addcontentsline{toc}{#1}{\ifnum #2>\c@secnumdepth \else
        \protect\numberline{\csname the#1\endcsname}\fi #7}}%
  \fi \@xsect{#5}}

\long\def\@makefigurecaption#1#2{\vskip 10mm #1. #2\par}

\long\def\@maketablecaption#1#2{\hbox to \hsize{\parbox[t]{\hsize}
  {#1 \\ #2}}\vskip 0.3ex}

\def\fnum@figure{Figure \thefigure}
\def\figure{\let\@makecaption\@makefigurecaption \@float{figure}}
\@namedef{figure*}{\let\@makecaption\@makefigurecaption \@dblfloat{figure}}

\def\table{\let\@makecaption\@maketablecaption \@float{table}}
\@namedef{table*}{\let\@makecaption\@maketablecaption \@dblfloat{table}}

\floatsep 10mm plus 4pt minus 4pt 
\textfloatsep=\floatsep           
\intextsep=\floatsep              

\long\def\@makefntext#1{\parindent 1em\noindent\hbox{${}^{\@thefnmark}$}#1}

\mathindent=0em

\def\maketitle{\begingroup        
    \def\thefootnote{\fnsymbol{footnote}}%
    \newpage \global\@topnum\z@
    \@maketitle \@thanks
  \endgroup
  \let\maketitle\relax \let\@maketitle\relax
  \gdef\@thanks{}\let\thanks\relax
  \gdef\@address{}\gdef\@author{}\gdef\@title{}\let\address\relax}

\def\justify@on{\let\\=\@normalcr
  \leftskip\z@ \@rightskip\z@ \rightskip\@rightskip}

\newbox\fm@box                    

\def\@maketitle{
  \global\setbox\fm@box=\vbox\bgroup
    \vskip 8mm                    
    \raggedright                  
    \hyphenpenalty\@M             
    {\Large \@title \par}         
    \vskip\@bls                   
    {\normalsize                  
     \@author \par}               
    \vskip\@bls                   
    \@address                     
  \egroup
  \twocolumn[
    \unvbox\fm@box                
    \vskip\@bls                   
    \unvbox\abstract@box          
    \vskip 2pc
\vskip -10mm]}                  

\newcounter{address}
\def\theaddress{\alph{address}}
\def\@makeadmark#1{\hbox{$^{\rm #1}$}}

\def\address#1{\addressmark\begingroup
  \xdef\@tempa{\theaddress}\let\\=\relax
  \def\protect{\noexpand\protect\noexpand}\xdef\@address{\@address
  \protect\addresstext{\@tempa}{#1}}\endgroup}
\def\@address{}

\def\addressmark{\stepcounter{address}%
  \xdef\@tempb{\theaddress}\@makeadmark{\@tempb}}

\def\addresstext#1#2{\leavevmode \begingroup
  \raggedright \hyphenpenalty\@M \@makeadmark{#1}#2\par \endgroup
  \vskip\@bls}

\newbox\abstract@box              

\def\abstract{%
  \global\setbox\abstract@box=\vbox\bgroup
  \small\rm
  \ignorespaces}
\def\endabstract{\par \egroup}

\def\thebibliography#1{\section*{REFERENCES}\list{\arabic{enumi}.}
  {\settowidth\labelwidth{#1.}\leftmargin=1.67em
   \labelsep\leftmargin \advance\labelsep-\labelwidth
   \itemsep\z@ \parsep\z@
   \usecounter{enumi}}\def\makelabel##1{\rlap{##1}\hss}%
   \def\newblock{\hskip 0.11em plus 0.33em minus -0.07em}
   \sloppy \clubpenalty=4000 \widowpenalty=4000 \sfcode`\.=1000\relax}

\newcount\@tempcntc
\def\@citex[#1]#2{\if@filesw\immediate\write\@auxout{\string\citation{#2}}\fi
  \@tempcnta\z@\@tempcntb\m@ne\def\@citea{}\@cite{\@for\@citeb:=#2\do
    {\@ifundefined
       {b@\@citeb}{\@citeo\@tempcntb\m@ne\@citea
        \def\@citea{,\penalty\@m\ }{\bf ?}\@warning
       {Citation `\@citeb' on page \thepage \space undefined}}%
    {\setbox\z@\hbox{\global\@tempcntc0\csname b@\@citeb\endcsname\relax}%
     \ifnum\@tempcntc=\z@ \@citeo\@tempcntb\m@ne
       \@citea\def\@citea{,\penalty\@m}
       \hbox{\csname b@\@citeb\endcsname}%
     \else
      \advance\@tempcntb\@ne
      \ifnum\@tempcntb=\@tempcntc
      \else\advance\@tempcntb\m@ne\@citeo
      \@tempcnta\@tempcntc\@tempcntb\@tempcntc\fi\fi}}\@citeo}{#1}}

\def\@citeo{\ifnum\@tempcnta>\@tempcntb\else\@citea
  \def\@citea{,\penalty\@m}%
  \ifnum\@tempcnta=\@tempcntb\the\@tempcnta\else
   {\advance\@tempcnta\@ne\ifnum\@tempcnta=\@tempcntb \else
\def\@citea{--}\fi
    \advance\@tempcnta\m@ne\the\@tempcnta\@citea\the\@tempcntb}\fi\fi}

\def\ps@crcplain{\let\@mkboth\@gobbletwo
     \def\@oddhead{\reset@font{\sl\rightmark}\hfil \rm\thepage}%
     \def\@evenhead{\reset@font\rm \thepage\hfil\sl\leftmark}%
     \let\@oddfoot\@empty
     \let\@evenfoot\@oddfoot}

\sloppy                         
\emergencystretch=1pc           
\flushbottom                    
\ps@crcplain                    
\newcommand{\ttbs}{\char'134}
\newcommand{\AmS}{{\protect\the\textfont2
  A\kern-.1667em\lower.5ex\hbox{M}\kern-.125emS}}

\hyphenation{author another created financial paper re-commend-ed}

\title{String Model Building in the Age of D-Branes}

\author{Joseph D. Lykken\address{Theory Group, MS106,\\
        Fermi National Accelerator Laboratory, \\
        P.O. Box 500, Batavia, IL 60510}
	\thanks{Talk given at the 5th International Workshop on
Supersymmetry and Unification of Fundamental Interactions (SUSY-96),
University of Maryland, College Park, May 29 - June 1, 1996.}
}


\begin{abstract}
The string duality revolution calls into question virtually
all of the working assumptions of string model builders.
A number of difficult questions arise. I use fractional
charge as an example of a criterion which one would hope
is robust beyond the weak coupling heterotic limit.
\end{abstract}

\maketitle

\section{The String Duality Revolution}
Recent developments in string theory have radically altered
our view of what superstrings are and what properties or
symmetries are fundamental in the nonperturbative regime.
The heterotic, Type I, Type IIA, and Type IIB superstrings
now appear to be different weak coupling limits of the same
underlying nonperturbative theory. Furthermore there is strong
evidence that this theory has at least one more well-defined
limit, called ``M theory'', which is not a weak coupling limit
and does not admit a world-sheet description. In addition the
existence of D-brane backgrounds in the Type I and II strings
implies that we will have to go beyond the world-sheet
description to correctly incorporate the effects of dynamical
D-branes.

These developments must be characterized as revolutionary since
they call into question the very idea that string theory is a
theory of ``strings''. At this point the world-sheet and
world-sheet symmetries seem less fundamental in string theory
than string duality symmetries and ``global symmetries'' like
spacetime supersymmetry. This paradigm shift holds real promise
for giving us at long last a tangible hold on nonperturbative
string dynamics at some point in the not-too-distant future.
However, as with most revolutions, we must
expect some strong doses of pain and confusion during the interregnum.

\section{Assumptions of String Model Builders}

For those in the already difficult business of
spinning gossamer threads between string theory and
particle physics, we are left with the disturbing fact that
the current revolution calls into question virtually
{\bf all} of the working assumptions of string model
builders. Without being exhaustive, let me gather
these assumptions into three groups:

\noindent $\bullet$ The heterotic/conformal field theory assumptions.

\noindent $\bullet$ The assumptions about relations between scales
and couplings.

\noindent $\bullet$ The assumptions about how gauge groups are
realized.

Let us consider in turn some of the questions which arise
with respect to these assumptions.

\subsection{Whither heteroticity?}

All of the promising attempts at connecting string theory
to the standard model take as their starting point the
weakly-coupled 10 dimensional heterotic string. In this limit
string theory was assumed to be rigorously described by
modular invariant conformal field theory.

However we know now that there are special regions in moduli
space which exhibit singular behavior associated with stringy
solitons becoming massless. This behavior is simply absent in
the conformal field theory description, even though it occurs
at arbitrarily weak coupling. Thus we are forced to conclude
that, {\bf even in the weak coupling heterotic limit}, conformal field
theory may give an incomplete description of essential physics
like the massless spectrum.

\noindent Q: Can we somehow generalize conformal field theory
to handle these singular behaviors?

\noindent Q: Does it matter? Are these behaviors mere curiosities,
or do they infect the phenomenologically important regions of
moduli space?

Model builders have focused on the heterotic string because
it did not appear possible to embed the standard model gauge
group and chiral fermions into the Type I or Type II string
compactifications. However these arguments are now known to
be invalidated by the presence of D-branes.

\noindent Q: Was the no-go theorem \cite{nogo}
for the Type II string
premature? Can we embed the standard model into the weak
coupling limit of Type II?

We must worry, moreover, whether it makes sense to extract
phenomenology from {\bf any} weak coupling limit of the string.
We already have a strong argument that stabilizing the dilaton
vev requires that we are in a region of moduli space with no
weak coupling interpretation. In fact the only compelling
reason to believe that weak coupling string theory has {\bf any}
connection to the real string ground state is the
phenomenological success of heterotic string models in
reproducing standard model physics. This makes it seem likely
that exact string symmetries protect the important features of
these models as we move in moduli space from the weak coupling
limit towards the real ground state.

Still, it is not premature to wonder about the phenomenological
properties of M-theory:

\noindent Q: Can we embed the standard model gauge group
and three generations of standard model fermions into
compactifications of M-theory?

\noindent Q: If there are ``realistic'' models in several
different limits of string theory, how do they map into
each other?

This latter question could be of key importance, if it turns
out that there are trajectories of realistic string vacua
that flow from one string limit to another.

\subsection{Scales and couplings}

String model building has heretofore assumed the relations
between scales and couplings derived in the perturbative
heterotic string. Thus, the effective gauge couplings are
assumed to have the form \cite{kl}:

\begin{equation}
{1\over g_a^2(p)} = {k_a\over g_{\rm string}^2}
+ {b_a\over 16\pi ^2}{\rm log}{M_{\rm string}^2\over p^2}
+ {\Delta _a\over 16\pi ^2}
\end{equation}
\noindent where $b_a$ are one-loop beta function coefficients
and $\Delta _a$ are model-dependent threshold corrections.
The Kac-Moody levels $k_a$ are positive integers, except for
the $U(1)$ hypercharge parameter $k_1$, which is a
continuous (real) free parameter $\ge$$1$.
The effective gauge coupling unification scale $M_{\rm string}$
is determined by the dilaton dependence of the heterotic
one-loop effects:

\begin{equation}
M_{\rm string} \sim g_{\rm string}M_{\rm Planck}
\end{equation}

The assumption that the 10 dimensional string coupling is not
strong restricts the compactification scale/scales to be not
too different from the string scale.

The bottom line is that in the standard scenario there
is basically one high scale: $M_{\rm string}$$\simeq$
$5$$\times$$10^{17}$ GeV. Above this scale we have stringy
effects and quantum gravity. Below this scale we have an effective
unified gauge theory. Gauge coupling unification occurs
somewhere between the string scale and (if there are large
threshold effects) the LEP preferred scale $3$$\times$$10^{16}$
GeV.

This simple picture has inspired many fruitful collaborations
between field theory and string theory model builders.
Each group is instructed to do their own thing, and then
attempt a graceful link-up with the other side over the
narrow no-man's land between $3$$\times$$10^{16}$ and
$5$$\times$$10^{17}$ GeV.

\noindent Q: Is this simple picture obsolete?

At least on the surface it appears that the relationships
between scales and couplings in string theory are
highly dependent on where you are in moduli space.
The relationships above which hold for the weakly-coupled
heterotic string are modified for the other weak coupling
limits of the string. In the M-theory limit it has been
suggested that there may not even be a simple transition
from a stringy regime to an effective 4 dimensional
field theory. Instead gravity may see an effective
5th dimension turn on around $10^{15}$ GeV, while the
gauge interactions remain effectively 4 dimensional and
unify at $3$$\times$$10^{16}$ GeV \cite{banks}.

As of now, it is difficult to put any meaningful limits
on what can occur in the murky depths of the strong
and intermediate coupling region of string theory.
To the extent that the theory is essentially quantum
in nature, geometrical reasoning is likely to be unsound.
Thus, for example, one cannot argue that a large hierarchy
of scales or couplings necessarily corresponds to some
compactification radius becoming large. It is possible
that inherently stringy physics is not confined to scales
near $M_{\rm Planck}$, and may even be lurking at scales
within reach of future colliders \cite{wss}.

\noindent Q: What are the physically distinct scales of
string theory, and how tight are the relationships
between them?

\subsection{Gauge groups}

String model builders have assumed that gauge fields in
string models arise from the intrinsic $E_8$$\times$$E_8$
or $SO(32)$ gauge groups of the 10 dimensional heterotic
string, plus the Kaluza-Klein gauge fields arising from
compactification. This implies that the gauge group of
the effective 4 dimensional gauge theory has rank $\le$$22$,
and is restricted in other ways which become more and more
restrictive for higher Kac-Moody levels $k_a$.

This rank restriction is now known to be incorrect
at certain points in moduli space, even for the weak-coupling
limit of the heterotic string \cite{si}.
Additional gauge bosons
can arise as massless solitons of string theory.
For example, in a 6 dimensional $K3$ compactification
of the $SO(32)$ heterotic string (which can obviously be
compactified further to produce 4 dimensional examples),
the compactification requires a nontrivial gauge background
with instanton number 24. There is a point in moduli space
where the size of all 24 instantons shrinks to zero, and
at this point the rank of the gauge group increases by 24:

\begin{equation}
SO(32) \rightarrow SO(32) \times Sp(48)
\end{equation}

Conformal field theory misses this entirely -- there are
no conformal currents corresponding to these gauge fields.
Roughly speaking, the Kac-Moody level of this $Sp(48)$
is effectively zero.

\noindent Q: Can this happen in realistic models?

\noindent Q: Could the standard model gauge bosons be
massless solitions as viewed from the heterotic limit?
Is this consistent with the existence of the standard
model chiral fermions?

\noindent Q: Can the physical information encoded in
Kac-Moody levels be generalized beyond conformal
current algebras?

The latter question is important because in
modular invariant conformal field theory, the
Kac-Moody level restricts the massless spectrum of the
string model. Furthermore, gauge coupling unification
in the strict sense only occurs (at tree-level) if
$k_3/k_2 = 5k_3/3k_1 = 1$. Kac-Moody levels --in the
standard scenario-- have been a powerful tool for
classifying superstring phenomenology.

\section{Fractional Charge}

Superstring phenomenology has so far produced
few predictions, and very few of these could be
meaningfully called generic. The existing examples
of string models which embed the standard model gauge
group and precisely three standard model generations
do have some important features in common: they have
a hidden sector, the possibility of extra $U(1)$ gauge
factors, and highly restricted superpotential couplings.

An appealing generic prediction of string pheneomenology
was Schellekens' theorem \cite{schell},
which states that under a broad
and specific set of circumstances string models must
contain particles which violate the standard model charge
quantization condition (e.g. colorless particles with
electric charge 1/2, color triplets with charge 1/6, etc.).
Since the lightest fractionally charged particle is of
necessity stable, this is an interesting prediction even
if these particles are very heavy.

Briefly, Schellekens' theorem states that in any string
model which contains the standard model gauge group
{\bf one} of the following must be true:

\noindent 1. There are particles with fractional electric
charge.

\noindent 2. The model has $SU(5)$ unbroken at the string scale.

\noindent 3. The standard model gauge group is realized
at Kac-Moody levels greater than one.

Unfortunately, this theorem is proved using conformal
field theory and modular invariance of the weakly-coupled
heterotic string. Thus one of the most interesting generic
results of string phenomenology is now in doubt.

\noindent Q: Can we say anything about the possibility
or necessity of fractional charge beyond the weak coupling
heterotic limit?

\section{Fractional charge and Kac-Moody levels}

I have argued above that both Kac-Moody levels and Schellekens'
theorem have been useful tools of superstring
phenomenology, tools which we should try hard to generalize
in the ``new'' string theory. To emphasize this point I
will close by classifying some (most) of the existing
{\bf three generation} string models along these lines.

\noindent 1. Models with $SU(3)_c$$\times$
$SU(2)_L$$\times$$U(1)_Y$ at the string scale, with
$k_3$$=$$k_2$$=$$1$, $k_1$$=$$5/3$:

\noindent The known examples are the various fermionic models of
Faraggi \cite{far}.
Schellekens' theorem says there must be
fractional charge, and indeed there are fractionally charged
states in the massless spectrum. These exotics are
vectorlike, and some or all of them may get superheavy
masses by coupling to singlet vevs.

\noindent 2. Models with $SU(3)_c$$\times$
$SU(2)_L$$\times$$U(1)_Y$ at the string scale, with
$k_3$$=$$k_2$$=$$1$, $k_1$$\ne$$5/3$:

\noindent The known examples are the $Z_3$ and $Z_3$$\times$$Z_3$
(0,2) abelian orbifolds \cite{fiqs89},
as well as the fermionic models of Chaudhuri et al \cite{chl}.
These models have fractionally charged exotics as above.

\noindent 3. Models which embed
$SU(3)_c$$\times$$SU(2)_L$$\times$$U(1)_Y$ into a larger
group at the string scale (subsequently broken by Higgs vevs),
with $k_3$$=$$k_2$$=$$1$:

\noindent There are two cases. The gauge group can be
$SU(5)$$\times$$SU(5)$,
in which case we have the models of
Finnell and Maslikov et al \cite{fin}.
Because of the unbroken $SU(5)$, these models need not have
fractionally charged particles, and indeed they do not.
The only other way to have a GUT-like unified group with
$k_3$$=$$k_2$$=$$1$ is the ``flipped $SU(5)$''
fermionic model \cite{flip}.
Because this is {\bf not} standard $SU(5)$ there are
fractionally charged exotics. However flipped $SU(5)$ has
the nice property that these exotics are also 4's and
$\bar 4$'s of a hidden $SU(4)$; thus fractional electric
charge is confined.

\noindent 4. Models which embed
$SU(3)_c$$\times$$SU(2)_L$$\times$$U(1)_Y$ into a larger
group at the string scale (subsequently broken by Higgs vevs),
with $k_3$$=$$k_2$$>$$1$:

\noindent Until recently the only examples were uninteresting
because they had chiral color sextet exotics.
However there is now an example of $SO(10)$ model with
Kac-Moody level three \cite{tye}. There is no fractional charge.

\noindent 5. Models with $SU(3)_c$$\times$
$SU(2)_L$$\times$$U(1)_Y$ at the string scale, with
$k_3$$=$$k_2$$>$$1$:

\noindent This case is interesting because it
employs the second of the two loopholes in
Schellekens' theorem.
Until recently there were no known examples.
However one can argue that the existence of
the Finnell model also implies the existence
of models of this type. A detailed examination of the
superpotential in the Finnell model indicates
that the $(5,\bar 5)$ and $\bar 5,5)$ Higgs fields
are probably moduli. In such a case Higgs breaking
is equivalent to continuous Wilson line breaking,
indicating that there are
$SU(3)_c$$\times$$SU(2)_L$$\times$$U(1)_Y$ string
vacua with $k_3$$=$$k_2$$=$$2$, $k_1$$=$$10/3$.

In fact, we now have a direct fermionic construction
of a three generation model of this type \cite{us}.
The full gauge group is

\begin{eqnarray*}
\lefteqn{\big[ SU(3)_c\times SU(2)_L\times U(1)_Y \big]
\times } \\
\lefteqn{\big[ SU(2)\times SU(2)\big]_{\rm hidden}
\times \big[ U(1) \big]^7 \times U(1)_{\rm anomalous} }
\end{eqnarray*}
\noindent Some or all of the extra $U(1)$'s will be broken by the
Green-Schwarz mechanism which eliminates the anomalous
$U(1)$.

The complete massless matter spectrum, not including
standard model singlets, consists of the following
$N$$=$$1$ chiral superfields:

\noindent -- three standard model generations plus up and
down type Higgs.

\noindent -- five $(3,1)$$+$$(\bar 3,1)$ vectorlike pairs
of exotic charge $\pm$1/3 quarks.

\noindent -- eight pairs of charge $\pm$$1$ weak doublets.

\noindent -- two pairs of charge $\pm$$1$
$SU(3)_c$$\times$$SU(2)_L$ singlets.

\noindent As expected, there are no fractionally charged particles.

\section{Conclusion}

The phenomenology of the three generation weakly-coupled
heterotic string models is trying to tell us something
profound about particle physics. Clearly also the string
duality revolution is telling us something profound about
string theory. Somehow we must find a way to pay heed to
both messages.


\begin{thebibliography}{9}
\bibitem{nogo} L. Dixon, V. Kaplunovsky, Nucl. Phys. B294 (1987) 43.
\bibitem{kl} V. Kaplunovsky and J. Louis,
Nucl. Phys. B444 (1995) 191.
\bibitem{banks} T. Banks and M. Dine,
``Coupling and Scales in Strongly Coupled Heterotic String Theory'',
hep-th/9605136.
\bibitem{wss} J. Lykken, ``Weak Scale Superstrings'',
hep-th/9603133.
\bibitem{si} E. Witten, Nucl. Phys. B460 (1996) 541.
\bibitem{schell} A. Schellekens, Phys. Lett. B237 (1990) 363.
\bibitem{far} A. Faraggi, Phys. Lett. B278 (1992) 131;
Nucl. Phys. B387 (1992) 239; B403 (1993) 101; B407 (1993) 57;
Phys. Lett. B326 (1994) 62; B329 (1994) 208; B339 (1994) 223.
\bibitem{fiqs89} A. Font, L. Ibanez, F. Quevedo, A. Sierra,
Nucl. Phys. B331 (1990) 421.
\bibitem{chl} S. Chaudhuri, G. Hockney, J. Lykken,
``Three Generations in the Fermionic Construction'',
hep-th/9510241.
\bibitem{fin} D. Finnell, Phys. Rev. D53 (1996) 5781;
A. Maslikov, S. Sergeev, and G. Volkov,
Phys. Rev. D50 (1994) 7440; A. Maslikov, I. Naumov,
and G. Volkov, Int. J. Mod. Phys. A11 (1996) 1117.
\bibitem{flip} I. Antoniadis, J. Ellis, J. Hagelin, and
D. Nanopoulos, Phys. Lett. B231 (1989) 65; J. Lopez,
D. Nanopoulos, and K. Yuan, Nucl. Phys. B399 (1993) 654;
Phys. Rev. D50 (1994) 4060;
J. Lopez and D. Nanopoulos, Nucl. Phys. B338 (1989) 73;
Phys. Rev. Lett. 76 (1996) 1566.
\bibitem{tye} Z. Kakushadze and H. Tye, ``Three Family
SO(10) Grand Unification in String Theory'',
hep-th/9605221.
\bibitem{us} G. Hockney and J. Lykken, to appear.
\end{thebibliography}
\end{document}